\documentclass[prb,twocolumn,amsmath,amssymb]{revtex4-1}

\usepackage{amsfonts}
\usepackage{graphicx}% Include figure files
\usepackage{natbib,hyperref}
\usepackage{bm}

\usepackage[normalem]{ulem}
\usepackage{color}

\begin{document}

\author{Areg Ghazaryan}
\affiliation{IST Austria (Institute of Science and Technology Austria), Am Campus 1, 3400 Klosterneuburg, Austria}

\author{Mikhail Lemeshko}
\affiliation{IST Austria (Institute of Science and Technology Austria), Am Campus 1, 3400 Klosterneuburg, Austria}

\author{Artem G. Volosniev}
\affiliation{IST Austria (Institute of Science and Technology Austria), Am Campus 1, 3400 Klosterneuburg, Austria}

\title{Filtering Spins by Scattering from a Lattice of Point Magnets}

\begin{abstract}
Nature creates electrons with two values of the spin projection quantum number. In certain applications, it is important to filter electrons with one spin projection from the rest.
Such filtering is not trivial, since spin-dependent interactions are often weak, and cannot lead to any substantial effect.
Here we propose an efficient spin filter based upon scattering from a two-dimensional crystal, which is made of aligned point magnets. The polarization of the outgoing electron flux is controlled by the crystal, and reaches maximum at specific values of the parameters. In our scheme, polarization increase is accompanied by higher reflectivity of the crystal. High transmission is feasible in scattering from a quantum cavity made of two crystals. Our findings can be used for studies of low-energy spin-dependent scattering from two-dimensional ordered structures made of magnetic atoms or aligned chiral molecules.
\end{abstract}

\maketitle

\section{Introduction}

The quest for spin filters started directly after the discovery of spin. It turns out that for electrons (in contrast to atoms), this problem is not trivial, since the Lorentz force and the uncertainty principle render it difficult, if not impossible, to achieve spin polarizarion using magnetic fields alone~\cite{Mott1929,batelaan1997,garraway1999}. The quest continues even after a century of research and numerous proposals~\cite{Gilbert2000,koga2002,Ciuti2002,Zhou2004,Karimi2012,Grillo2013,Dellweg2017,Ahrens2017}. The applications of polarizers are quite diverse and span atomic, molecular, nuclear, and condensed-matter physics~\cite{Kessler2013,Prescott1978,Subashiev1999,heckel2008,Gay2009}. They are used to study magnetization dynamics~\cite{Vollmer2003,Suzuki2010} and in spin and angle resolved photoemission spectroscopy of topological materials~\cite{Dil2019}, to give just a few examples. 

At the present time, not only inorganic but also organic systems are being considered as possible spintronic devices~\cite{Sanvito2011}. Recent experiments show that electrons become spin polarized when passing through a molecular monolayer of chiral molecules (such as DNA, oligopeptides, helicene, etc.)~\cite{Gohler2011,Xie2011,Kettner2015, Mishra2013,Dor2013,Einati2015,Kiran2016,Kettner2018}. This property of chiral molecules is now called chiral induced spin selectivity (CISS), and its existence can lead to novel spin filters~\cite{Naaman2015,naaman2019_review}. The magnitude of polarization in CISS is quite high, yet the intensity of the outgoing flux is relatively low. Despite the seeming simplicity of the CISS experiments, the observed effect is an outstanding problem in theoretical physics. Several models that rely on scattering from a single molecule have been suggested  ~\cite{Yeganeh2009,Medina2012,Varela2013,Guo2012,Gutierrez2012,Gutierrez2013,Guo2014,Matityahu2016,Michaeli2015,Yang2019, Geyer2019,Gersten2013,Dalum2019,Fransson2019, Ghazaryan2020}. However, it is still not clear whether the effect can be observed at a single-molecule level or CISS requires electron scattering from multiple molecules. In particular, strong dependence of the asymmetry function on the doping level~\cite{Ray1999} suggests that multiple scattering might be important.

\begin{figure}
\includegraphics[width=1.0\columnwidth]{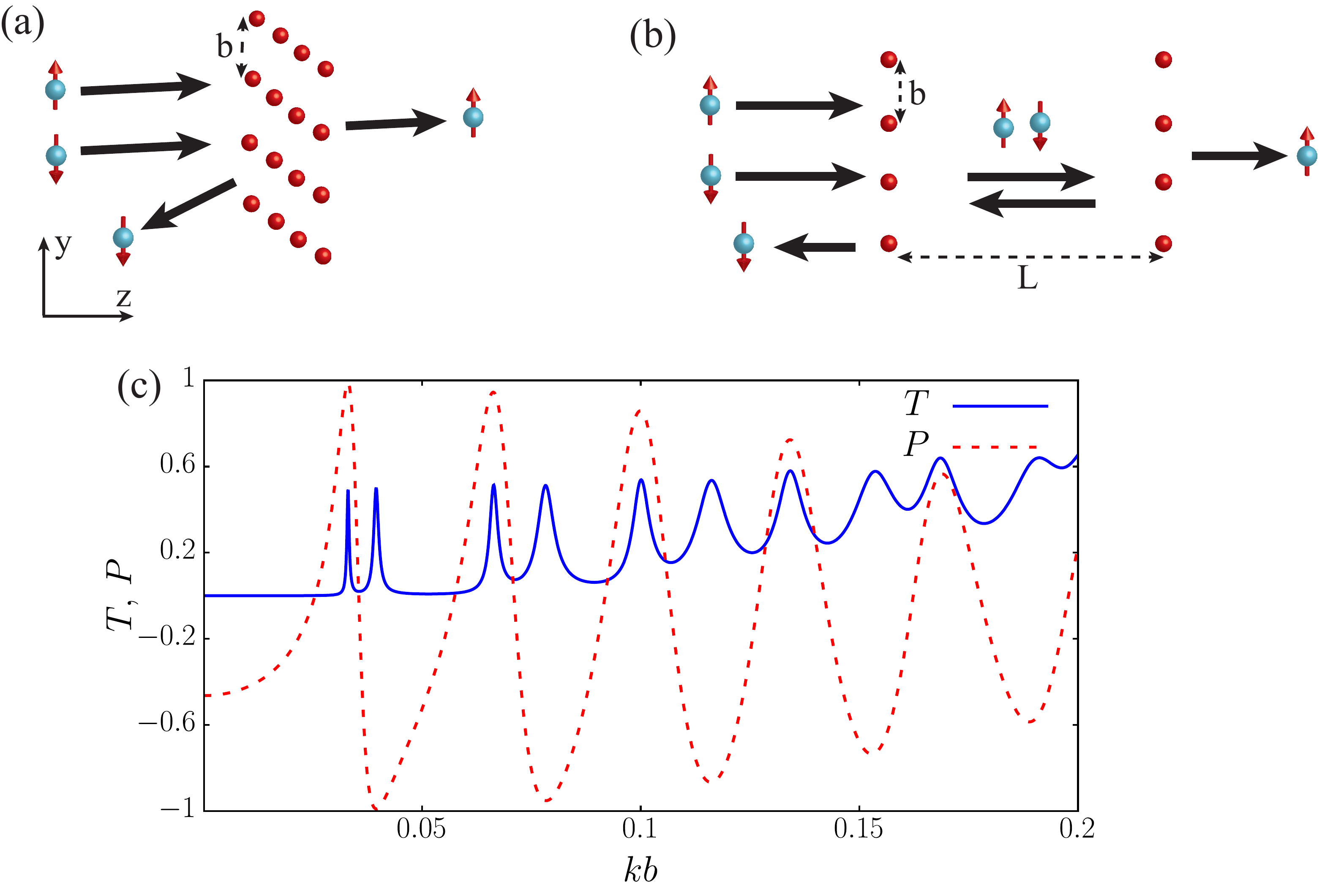}
\caption{\label{fig:PictorialFigure} Spin filter implementation in scattering from point magnets. (a) Unpolarized electrons scatter from a single layer of point magnets. The magnets form a square lattice with a lattice constant $b$. Close to the specific values of the parameters only `spin up' electrons are transmitted. (b) Unpolarized electrons scatter from two layers of point magnets. This set-up can enjoy both high polarization and transmission. Note that the electrons and magnets in (a) and (b) are depicted as finite-radius spheres for illustrative purposes --- in our theoretical model the magnets are pointlike, i.e., they are spheres with a vanishing radius. Our model is accurate for low-energy scattering~\cite{Braaten2006,Demkov2013}. (c) The transmission, $T$, and polarization, $P$, coefficients for scattering depicted in (b);  $T=\frac{T_{\uparrow}+T_{\downarrow}}{2}$, where $T_{s}$ is the transmission coefficient for a given projection of the spin. The parameters of the magnets are $a_0=0$, $a_1=0.02b$. The distance between the layers, $L$, is approximately $88b$.
\label{fig:1}
}\end{figure}

Our aim below is to investigate scattering from a two-dimensional (2D) layer of spatially arranged point scatterers (magnets), see Fig.~\ref{fig:1} (a). In the vicinity of the specific values of the parameters, the layer acts as a perfect mirror and can be used as a spin filter for low-energy electrons: while one spin component is perfectly reflected, the other one is transmitted. In addition, we analyze scattering from two layers of magnets, which can form a spin filter that enjoys both high transmission and polarization, see Figs.~\ref{fig:1} (b) and (c). Below, we discuss our results in more detail, and demonstrate that they give insight into CISS as a collective-scattering phenomenon.

\section{Results}

\subsection{Single Layer}

Our spin filter works with electrons impinging perpendicular to an infinite layer of magnets, see~Fig.~\ref{fig:1}~(a), which are modelled by contact potentials. 
The contact (also zero-range or pointlike) potential is a mathematical approximation~\cite{Bethe1935,Braaten2006, Demkov2013} to low-energy scattering. In our work, it amounts to replacing the electron-magnet interaction potential by a spin-dependent boundary condition at zero electron-magnet separation. If the parameters of the layer are tuned to specific values, the transfer of one spin component is impossible and the layer turns into a perfect spin filter. Our scheme thus realizes a spin filter with high polarization, which we discuss in detail below.

We start our theoretical description by considering scattering from a single layer, which bears some similarity to a 1D spin filter with a spin-dependent energy profile~\cite{Zhou2004}.  In order to illustrate the concept, we place the scatterers in the nodes of a lattice, i.e., at $\mathbf{a}_{lm}=lb\hat{\mathbf{x}}+mb\hat{\mathbf{y}}+0\hat{\mathbf{z}}$, where $l$ and $m$ are integers and $b$ is the lattice constant. The parameter $\mathbf{a}_{lm}$ determines the position where the `zero-range-potential' boundary condition should be enforced. Although, we have assumed that $\mathbf{a}_{lm}$ form a square lattice, we have checked that other geometries, e.g., a triangular lattice, lead to conceptually similar results. Finally, we assume that all scatterers have the same magnetization direction. This direction can be chosen arbitrarily in our theoretical analysis.

The parameters $\{\mathbf{a}_{lm}\}$ define a periodic structure, and, therefore, the electron wave function must be an eigenstate of a translation operator that shifts the wave function by $\mathbf{a}_{lm}$, i.e.,
$\Psi\left(\mathbf{r}+\mathbf{a}_{lm}\right)=e^{i\mathbf{k}_i\mathbf{a}_{lm}}\Psi\left(\mathbf{r}\right)$, where $\mathbf{k}_i$ is the momentum of an incoming electron. The corresponding scattering state reads as
\begin{equation}
\Psi\left(\mathbf{r}\right)=e^{ikz}+A\sum_{lm}\frac{e^{ik\left|\mathbf{r}-\mathbf{a}_{lm}\right|}}{\left|\mathbf{r}-\mathbf{a}_{lm}\right|},
\label{WaveFunc}
\end{equation}
where the incoming flux is given by the plane wave, and the outgoing flux is described by spherical waves propagating away from the point scatterers. We have assumed that $\mathbf{k}_i||\hat{\mathbf{z}}$ ($|\mathbf{k}_i|=k$), which implies that $\mathbf{k}_i\mathbf{a}_{lm}=0$.  
 The last term in Eq.~(\ref{WaveFunc}) is defined as the limit: $\lim_{R\to\infty}A_{R}\sum_{lm}^{R}$, $R$ is a dimensionless cut-off parameter, see the discussion below.
The constant $A_{R}$ is determined from the zero-range-potential boundary conditions~\cite{Demkov2013}:
\begin{equation}
\Psi\left(\mathbf{r}\rightarrow\mathbf{a}_{lm}\right)=s_{lm}\left(\frac{1}{\left|\mathbf{r}-\mathbf{a}_{lm}\right|}-\frac{1}{\alpha_{s}}\right),
\end{equation}
where $s_{lm}$ is the normalization constant, $\alpha_{s}$ is a spin-dependent scattering length that fully determines a zero-range potential. Note that our zero-range model describes low-energy scattering from potentials that decay faster than $1/r^3$ at large interparticle distances, provided that $b$ is larger than the range associated with the potential. In particular, our model is appropriate for electron-atom interactions ($\sim 1/r^4$ as $r\to \infty$).  
We assume that $\alpha_{\uparrow}=a_0+a_1$ and $\alpha_{\downarrow}=a_0-a_1$, where $\uparrow (\downarrow)$ denotes a spin projection of incoming electrons on the desired quantization axis;  $a_0$ ($a_1$) describes the spin-independent (spin-dependent) part of the potential. The quantization axis is chosen by the magnetization direction of the magnets, which, without loss of generality, in this work is given by the $y$-axis in Fig.~\ref{fig:1}~(a). Imposing the boundary condition, we obtain $A_R$:
\begin{equation}
A_R=-\alpha_s\left(1+\alpha_s\sum\limits_{\substack{lm \\ \mathbf{a}_{lm}\neq0}}^R\frac{e^{ik\left|\mathbf{a}_{lm}\right|}}{\left|\mathbf{a}_{lm}\right|}+ik\alpha_s\right)^{-1},
\label{AForm}
\end{equation}
where $R$ is used to define an upper limit of the sum. Equations~(\ref{WaveFunc}) and~(\ref{AForm}) fully determine all properties of scattering. 

To gain analytical insight, 
we explore the zero-energy limit ($k\rightarrow0$). The layer of magnets appears to be homogeneous for a distant observer ($|z|\gg b$), allowing us to focus on $\mathbf{r}=z\hat{\mathbf{z}}$. To write the wave function, we estimate the sums in Eqs.~(\ref{WaveFunc}) and~(\ref{AForm}) for large values of $R$ using the integral test: $\sum_{lm}^R\frac{1}{\left|\mathbf{r}-\mathbf{a}_{lm}\right|}\approx\frac{2\pi}{b^2}\left(\sqrt{R^2b^2+z^2}-|z|\right)$ and $\sum_{lm, \mathbf{a}_{lm}\neq0}^R\frac{1}{\left|\mathbf{a}_{lm}\right|}\approx\frac{2\pi}{b}\left(R-\Delta_0\right)$, where $\Delta_0\ge0$ is a constant, which depends only on the geometry of the system; it can easily be determined numerically, $\Delta_0\approx0.635$. Both sums diverge linearly with $R$ as $R\to\infty$, leading to a well-defined limit:
$\Psi\left(\mathbf{r}\right)=1+\frac{2\pi \alpha_s}{b^2}|z|-\frac{2\pi \alpha_s}{b}\Delta_0$,
which is identical to the 1D wave function that describes zero-energy scattering from the Dirac delta potential~\cite{Griffiths2018}, $g_s\delta(z)$: $\Psi_{1D}(z)=s\left(\frac{g_s}{2}|z|+1\right)$. This observation allows us to map the 3D problem onto a 1D zero-range model with
\begin{equation}
g_s=\frac{4\pi \alpha_s}{b(b-2\pi \alpha_s\Delta_0)}.
\label{PotAmplitude}
\end{equation} 
Considering finite-energy scattering from the potential $g_s\delta(z)$, we determine the transmission and reflection coefficients as $T_s=4k^2/\left(g_s^2+4k^2\right)$ and $R_s=g_s^2/\left(g_s^2+4k^2\right)$, respectively. The corresponding spin polarization is $P=\frac{T_\uparrow-T_\downarrow}{T_\uparrow+T_\downarrow}$. While Eq.~(\ref{PotAmplitude}) is accurate only for $k\to 0$, similar mapping exists also for finite values of $k$ (See Supplementary Note~1). It is clear from~Eq.~(\ref{PotAmplitude}), that the transmission coefficient vanishes when $b_c=2\pi \alpha_s\Delta_0$: quantum interference turns the layer of scatterers into a perfect mirror. 
Note that $g_{s}$ changes its sign at $b_c=2\pi \alpha_s\Delta_0$ implying the presence of a tightly bound state in the vicinity of $b_c=2\pi \alpha_s\Delta_0$. The state exists only in the proximity of $z=0$, which is outside the region of validity of the mapping onto 1D system. We have checked that this state does not appear in full calculations.

Let us analyze the polarization, $P$, in the two limiting cases: $a_0=0$ and $|a_1|\ll |a_0|$. There can be no polarization in scattering from a single zero-range potential with either $a_0=0$ or $a_1=0$. Therefore, the limits address the importance of multiple scatterings.  
For $a_0=0$, we derive 
$P=-16\pi^3a^3_1b\Delta_0 [k^2b^2\left(b^2-4\pi^2a^2_1\Delta_0^2\right)^2+4\pi^2a^2_1\left(b^2+4\pi^2a^2_1\Delta_0^2\right)]^{-1}$.
We can further simplify this expression assuming low-energy scattering, $kb^2\ll a_1$, and $a_1\ll b$: $P\approx-4\pi a_1\Delta_0/b$. In this limit $P\propto\sqrt{n}$, where $n=1/b^2$ is the density of scatterers. This dependence is a manifestation of coherent scattering, since for incoherent scattering one expects observables to be proportional to $n$. We find that $P\rightarrow 0$ for $b\rightarrow\infty$,  recovering the fact that a single scatterer with $a_0=0$ cannot act as a spin polarizer.  
In the other limit, $|a_1|\ll |a_0|$, we derive
$P\approx-8\pi^2a_0a_1b[k^2b^2\left(b-2\pi a_0\Delta_0\right)^3+4\pi^2a^2_0\left(b-2\pi a_0\Delta_0\right)]^{-1}$.
Taking again the limit of $kb^2\ll |a_0|$ and $|a_0|\ll b$, we obtain $P\approx-2a_1/a_0-4\pi a_1\Delta_0/b$, which has the same density dependence as the previous case, although, a single scatterer can now act as a weak spin polarizer -- the corresponding polarization is $P\approx-2a_1/a_0$. 
For electrons with $kb^2\sim a_0$, there is a competition between the two terms in the denominator of $P$, which makes the dependence on $n$ more complex. Finally, we note that the polarization reaches its maximum value in the vicinity of the specific lattice spacing, $b_c\approx2\pi a_0\Delta_0$; the transmission vanishes at the same time. This regime holds promise for constructing a spin filter, as we discuss below.

\begin{figure}
\includegraphics[width=1.0\columnwidth]{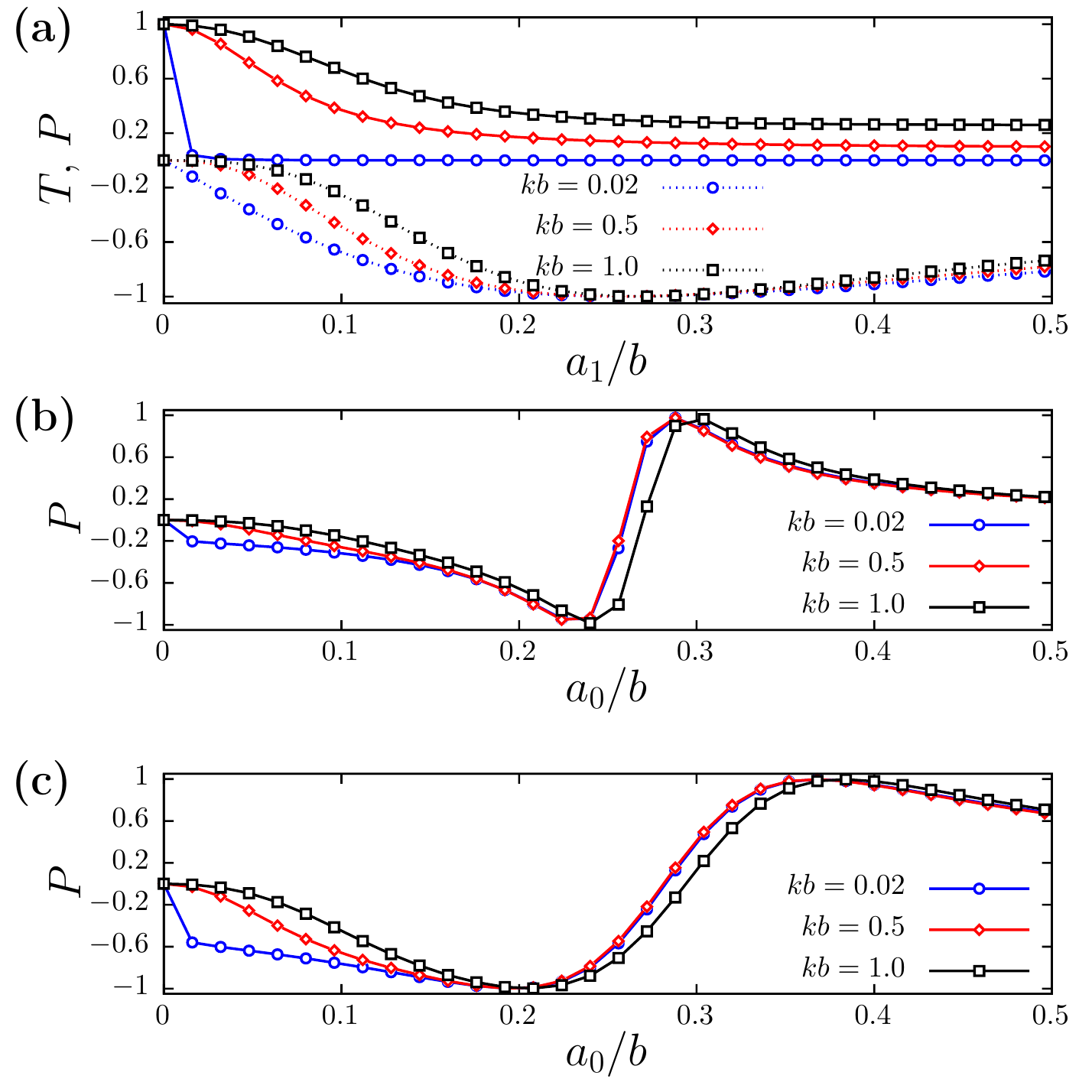}
\caption{\label{fig:TransPolSingleSheet} Transmission, $T$, and polarization, $P$, coefficients for scattering of electrons from a single 2D crystal. (a) Dependence of transmission (points connected by solid curves) and the polarization (points connected by dotted curves) on the dimensionless scattering length, $a_1/b$, when $a_0=0$. The transmission coefficient is defined as $T=\frac{T_{\uparrow}+T_{\downarrow}}{2}$, where $T_{s}$ is the transmission coefficient for a given projection of the spin. (b) Dependence of the polarization coefficient on $a_0/b$ when $a_1/a_0=0.1$. (c) Dependence of the polarization coefficient on $a_0/b$ when $a_1/a_0=0.3$.  
}\end{figure}

Having analyzed the zero-energy limit, we consider Equations~(\ref{WaveFunc}) and~(\ref{AForm}) for finite values of $k$. For low energies, we establish a 1D mapping similar to Eq.~(\ref{PotAmplitude}), see Supplementary Note~1. This mapping does not yield any qualitatively new results in comparison to  Eq.~(\ref{PotAmplitude}), as we illustrate below for a set of parameters. 
Figure~\ref{fig:TransPolSingleSheet}~(a) shows the dependence of transmission and polarization on the dimensionless scattering length $a_1/b$ for different values of electron momenta, assuming that $a_0=0$. As was described above, for the specific ratio of $a_1/b$ the transmission $T_\uparrow$ goes to zero and the layer of magnets acts as a perfect mirror for electrons with `spin up', which maximizes the polarization coefficient. While Eq.~(\ref{PotAmplitude})  implies that the position of zero transmission does not depend on $k$, that is no longer the case for the full solution. We do observe a minor change of the peak position in Fig.~\ref{fig:TransPolSingleSheet}~(a), as explained in detail in Supplementary Note~1. For small momenta, the transmission $T=\left(T_\uparrow+T_\downarrow\right)/2$ is vanishing everywhere in the region with noticeable polarization, however, the situation changes if $kb$ is increased. At $kb=1.0$ there is already a range of $a_1/b$ where both transmission and polarization are substantial.
\begin{figure}
\includegraphics[width=1.0\columnwidth]{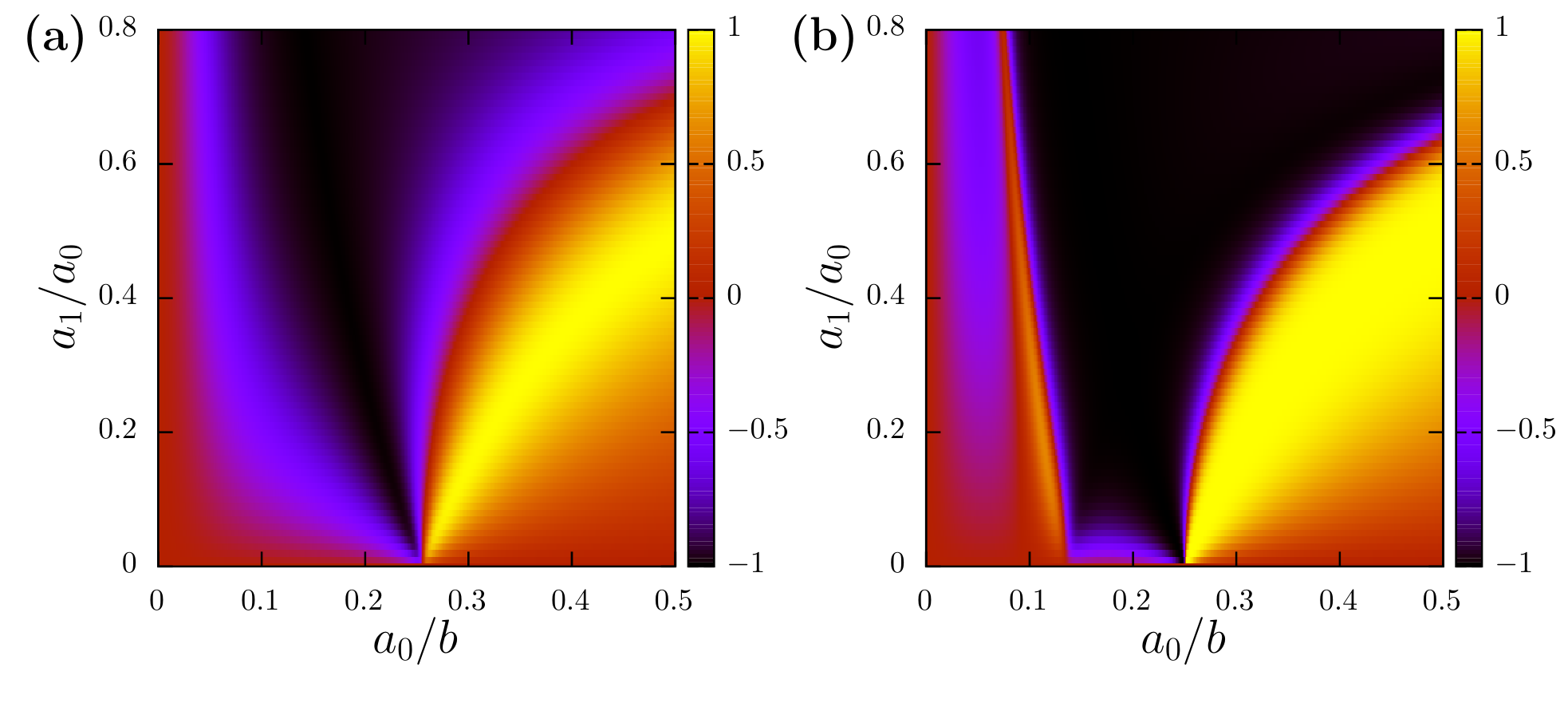}
\caption{\label{fig:PolDensityPlot.pdf} Polarization coefficient, $P$, for extended range of system parameters. The density plot of the polarization coefficient as a function of the dimensionless scattering lengths $a_0/b$ and $a_1/a_0$ for (a) one, and (b) two 2D crystals. The separation between the crystals is $L=100b$ in (b);  the momentum of electrons is $kb=0.5$ in both panels. 
}\end{figure} 
These results for $kb=1$ are accurate as long as $b\gg r_{\mathrm{eff}}$, where $r_{\mathrm{eff}}$ is the effective range. Indeed, a zero-range potential works only for small values of $k r_{\mathrm{eff}}$~\cite{Braaten2006}, otherwise electrons resolve the inner part of the interaction potential. 
 Figure~\ref{fig:TransPolSingleSheet} also shows the dependence of polarization on $a_0/b$ for $a_1/a_0=0.1$ (b) and $a_1/a_0=0.3$ (c). The value of $kb$ does not have any important effect on the position of the peak. Still, working with larger momenta is beneficial, since it modifies transmission considerably (similar to Fig.~\ref{fig:TransPolSingleSheet}~(a)). The sign change of the polarization coefficient presented in Figs.~\ref{fig:TransPolSingleSheet} (b) and (c) follows from Eq.~(\ref{PotAmplitude}). Assuming that both $\alpha_{\uparrow}$ and $\alpha_{\downarrow}$ are positive, the mirror (point where $T_{\uparrow}=0$) for spin-up electrons is at $b_c^{\uparrow}=2\pi \alpha_{\uparrow}$, and it is at $b_c^{\downarrow}=2\pi \alpha_{\downarrow}$ for spin-down electrons. The former (latter) mirror leads to negative (positive) polarization. Somewhere, in between these two points polarization must vanish, which leads to points with zero polarization in Figs.~\ref{fig:TransPolSingleSheet} (b) and (c).  The width of the region with high polarization in Fig.~\ref{fig:TransPolSingleSheet} (c) is larger than that in Fig.~\ref{fig:TransPolSingleSheet} (b). This width is controlled by the ratio $a_1/a_0$, as we demonstrate in Fig.~\ref{fig:PolDensityPlot.pdf} (a), which presents the density plot of polarization as a function of $a_0/b$ and $a_1/a_0$.  

\begin{figure}
\includegraphics[width=1.0\columnwidth]{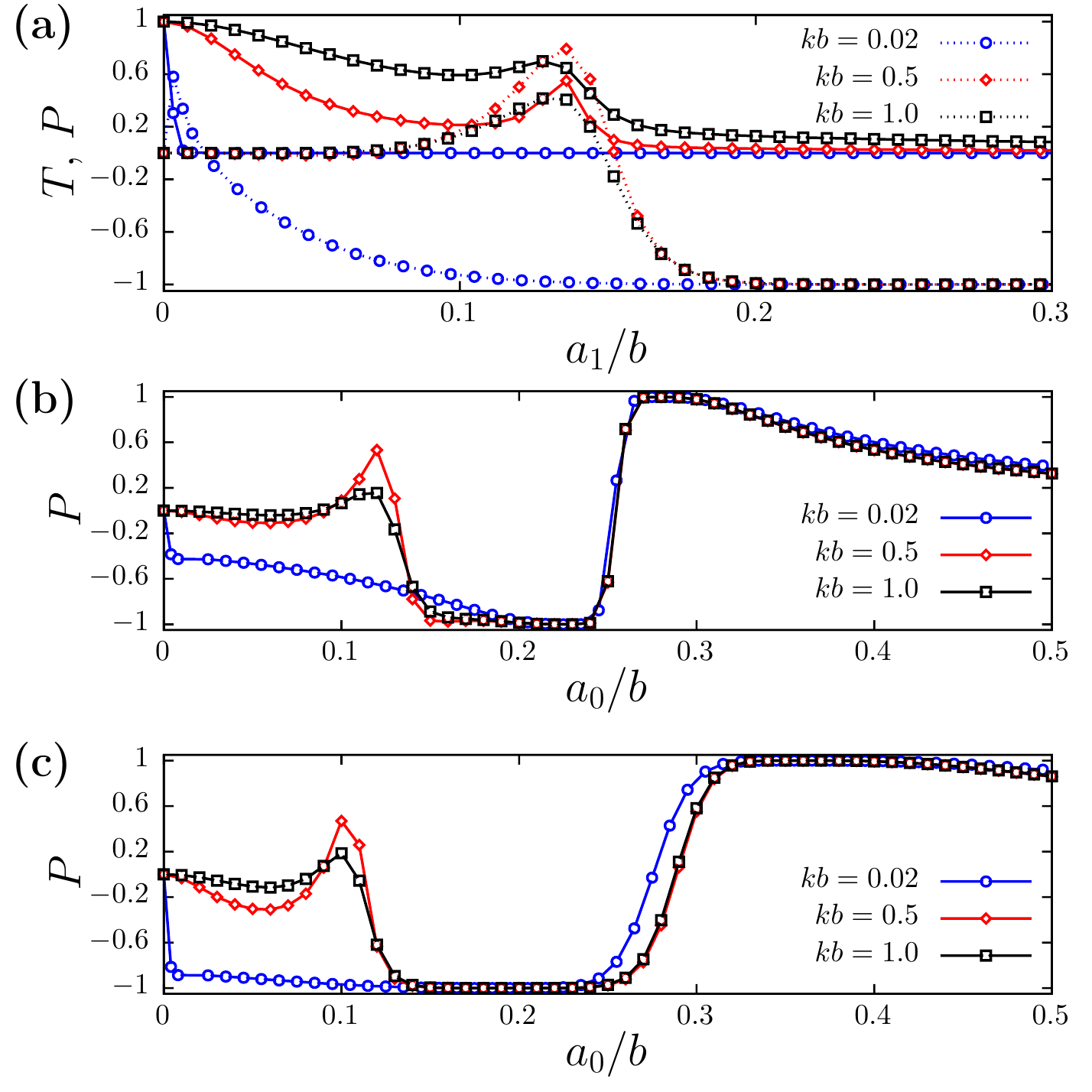}
\caption{\label{fig:TransPolDoubleSheet} Transmission, $T$, and polarization, $P$, coefficients for scattering of electrons from two 2D crystals. (a) Dependence of transmission (points connected by solid curves) and the polarization (points connected by dotted curves) on the dimensionless scattering length, $a_1/b$, when $a_0=0$. The transmission coefficient is defined as $T=\frac{T_{\uparrow}+T_{\downarrow}}{2}$, where $T_{s}$ is the transmission coefficient for a given projection of the spin. (b) Dependence of the polarization coefficient on $a_0/b$ when $a_1/a_0=0.1$. (c) Dependence of the polarization coefficient on $a_0/b$ when $a_1/a_0=0.3$.  The separation between the crystals is given by $L=100b$. 
}\end{figure}

\subsection{Two Layers} Figure~\ref{fig:TransPolSingleSheet}~(a) shows the transmission-polarization tradeoff present in our set-up: larger values of $P$ lead to smaller values of $T$.  To overcome this problem, we consider two aligned identical 2D sheets, see Fig.~\ref{fig:1}~(b). The sheets form a resonating cavity in the vicinity of the specific point, which can be used to tune scattering properties. We demonstrate this in Fig.~\ref{fig:1}~(c), where the peaks in transmission due to the internal levels of the cavity are accompanied by the peaks of the polarization. The latter peaks are connected to the ones present in scattering from a single layer, although their position is no longer predicted by Eq.~(\ref{PotAmplitude}).  Figure~\ref{fig:1}~(c) confirms that one can engineer an efficient spin filter with high transmission and polarization for a given initial wave packet with a well defined peak at a certain (low-energy) value of $k$.

Similar to the single-layer case, we cast the 3D problem onto the 1D model with two delta-function potentials of the strength~(\ref{PotAmplitude}), see Supplementary Note~2. This approximation is accurate as long as $L\gg b$, where $L$ is the distance between the potentials (layers). Naturally, transmission and polarization depend on $L$. The value used in Fig.~\ref{fig:1}~(c) was chosen to maximize $T$ and $P$ for the first peak. To demonstrate that a quantum cavity can increase transmission for a general value of $L$, in this section we take $L=100b$, which has no special meaning in our problem.
Since for a single layer the results were (almost) energy-independent, we work below with the zero-energy mapping of Eq.~(\ref{PotAmplitude}), whose accuracy for two layers is validated in~Supplementary Note~2.
The transmission coefficient for scattering from two zero-range potentials is $T_s=\left|4k^2/\left[g_s^2e^{2ikL}+\left(ig_s+2k\right)^2\right]\right|^2$.  We do not present the expression for polarization -- it is cumbersome and does not provide us with any further insight. Instead, we analyze scattering for the parameters used in Fig.~\ref{fig:TransPolSingleSheet} to illustrate the single-layer case.  
Figures~\ref{fig:PolDensityPlot.pdf}~(b) and~\ref{fig:TransPolDoubleSheet}~(a)-(c) present the transmission and polarization coefficients for scattering from two layers.  Interference inside the cavity leads to additional peaks for both transmission and polarization.  These peaks can be used to engineer regions where both polarization and transmission are substantial as in Fig.~\ref{fig:1} (c). Our conclusion is that two sheets of quantum scatterers have enough tunability to allow for an efficient spin filter. The fact that the inter-sheet separation can be several orders of magnitude larger than the spacing between quantum scatterers makes it feasible to engineer such a filter with GaAs superlattices as we briefly outline below.
In this subsection, we have assumed that there is no attenuation of the electron current, and that electrons move balistically in between the layers. As will be shown below, these assumptions can be accurate for current experimental techniques.

\section{Discussion}

To summarize, we have shown that quantum interference in scattering from a 2D crystal can lead to spin filtering. In our model, a layer of spatially arranged point scatterers (magnets) at a specific point acts as a perfect mirror for one spin component, but still transmits electrons with another spin component.  Even though a single-layer spin filter suffers from a reflection-polarization tradeoff, we have demonstrated that two parallel sheets of scatterers can provide simultaneously high transmission and high polarization.
It makes sense to introduce some energy dependence, a potential, in between the layers, either to reflect some material or as an additional tuning parameter for quantum simulations~\cite{Lebrat2019, corman2019}. Spin filters with desired transmission and reflection coefficients are then obtained by global searching the space of all possible potentials and values of $L$~\cite{smith2019}. We leave this investigation to future studies, as we do not expect a slow-varying potential to change qualitatively our findings.  A complex potential can account for attenuation (absorption) of the electron current in between the layers~\cite{Patrici1996}. This effect should be small for a reasonably pure sample, and we do not consider it here.

A possible experimental realization of the suggested spin filter is to dope the outer layer of a GaAs superlattice with two layers of magnetic adatoms. This should be possible without considerable fine tuning, because current state-of-the-art polarizers for microscopy applications are already based on negative electron affinity GaAs superlattice photocathodes~\cite{pierce1975,Kuwahara2012,Liu2016,Cultrera2020}. The observed spin polarization in those set-ups can be larger than 80\% and the corresponding quantum efficiency is on the order of several percent.

\begin{figure}
	\centering
\includegraphics[width=0.6\columnwidth]{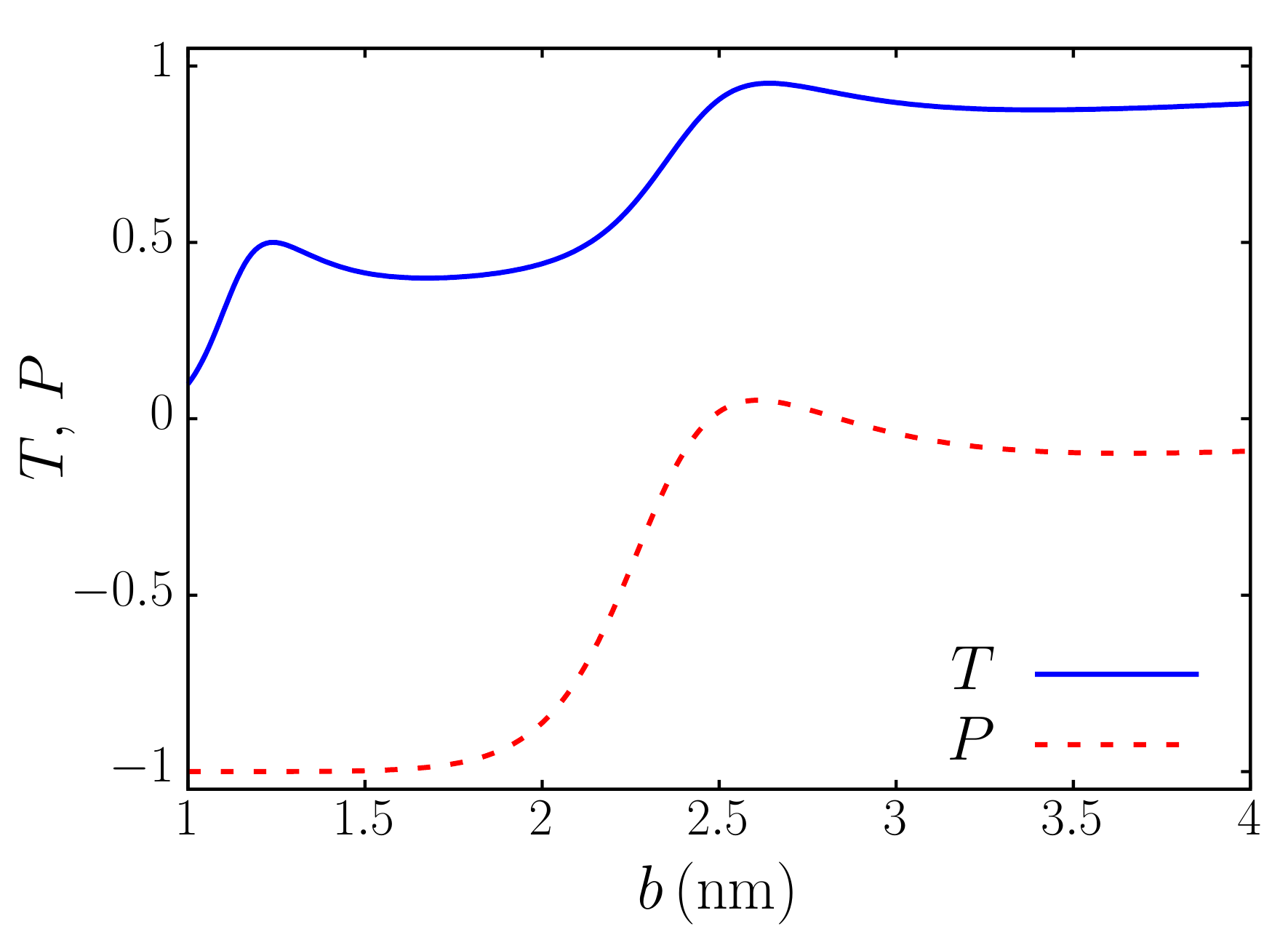}
\caption{\label{fig:TransPolDoubleGaAs} Transmission, $T$, and polarization, $P$, coefficients for scattering of electrons from two 2D crystals formed by Mn atoms on a GaAs sample. The coefficients are presented  as functions of the inter-Mn separation, $b$. The transmission coefficient is defined as $T=\frac{T_{\uparrow}+T_{\downarrow}}{2}$, where $T_{s}$ is the transmission coefficient for a given projection of the spin. The parameters that determine properties of the magnets are $a_0=0.12 \, \mathrm{nm} $, and $a_1=0.07\,\mathrm{nm}$. The separation between the layers is given by $L=80\,\mathrm{nm}$. The momentum of the electrons is $k\approx0.19\,\mathrm{nm^{-1}}$ (corresponding to the energy $20\,\mathrm{meV}$).
}\end{figure}

To justify these claims, we consider a GaAs superlattice doped with two layers of manganese (Mn) atoms. Mn atoms can be considered as point magnets for our purposes, since there are negligible spin-flip effects in low-energy $e^-+$ Mn scattering~\cite{Meintrup2000}. To find $a_0$ and $a_1$ we use the existing theoretical calculations on scattering cross sections~\cite{Dolmatov2013}, and estimate $a_0=(\sqrt{\sigma_{\downarrow}}+\sqrt{\sigma_{\uparrow}})/(4\sqrt{\pi})$ and  $a_1=(\sqrt{\sigma_{\uparrow}}-\sqrt{\sigma_{\downarrow}})/(4\sqrt{\pi})$,
where $\sigma_{s}$ is the total elastic-scattering cross-section for zero-energy spin-$s$ electrons; the quantization direction here is given by the spins of the electrons in the semi-filled shell of Mn atoms. For the sake of discussion, we use the SPRPAE2 data, which leads to $a_0\simeq 0.12 \, \mathrm{nm} $ and $a_1 \simeq 0.07 \,\mathrm{nm}$. In order to get transmission enhancement by two layers, electron propagation between the layers should be ballistic. Therefore, we consider the separation of the layers $L=80\,\mathrm{nm}$, which is comparable to the mean free path of the electrons in GaAs/AlGaAs superlattices \cite{Rauch1999} and considerably smaller compared to pure GaAs samples \cite{Brill1996}. We take the electron energy to be $20\,\mathrm{meV}$ for which case $kr_\mathrm{eff}\ll1$ ($r_\mathrm{eff}$ is estimated using the Van der Waals length $\sim 0.2\, \mathrm{nm}$), allowing us to apply the theory developed in this paper. Figure~\ref{fig:TransPolDoubleGaAs} shows that considerable polarization and transmission is obtained for $b\sim 1.5\,\mathrm{nm}$. This value of inter-Mn separation is not drastically different from $4\,\mathrm{nm}$ observed in the experiment \cite{Prucnal2015}. Therefore, we expect that effects considered in the current paper are within reach with currently available experimental techniques.

Our findings are connected to light scattering from an array of point dipoles~\cite{Abajo2007,Bordag2015} (although the latter has an additional complication due to the polarization of light). In particular, cooperative resonances in light scattering allow for a regime where a sheet acts a perfect mirror~\cite{Bettles2016,Shahmoon2017}, which is similar to what we find in our model. Our work adds another degree of freedom (spin) to this discussion, and acknowledges spin-filtering capabilities of a layer of point scatterers.

Our ideas do not employ fundamental properties of electrons, and can be used to implement spin filters in other systems as well.  Our proposal could be tested with cold atoms -- a tunable testground for studying quantum transport phenomena~\cite{chien2015}. Layers of atoms created with optical lattices~\cite{bloch2008} could simulate point magnets. Another type of atoms would then be used to simulate electrons, in particular, electron's spin would be modelled by a hyperfine state of the atom. For example, one could use $^{87}$Rb to simulate electrons, and an optical lattice loaded with $^{40}$K atoms to simulate the 2D crystal~\cite{Thywissen2007}. Our work explores the regime $k_{th}b \simeq 0 - 1$, where the thermal de Broglie wave-vector reads $k_{th}=\sqrt{2\pi m(^{87}\mathrm{Rb})k_B T}/\hbar$ ($k_{B}$ is the Boltzmann constant). Assuming that $b\sim 1\mu$m, this regime maps onto temperatures from zero to a few $n$K, which is within reach of current experimental set-ups~\cite{Kasevich2015}. Note however that for cold-atom set-ups our model is accurate also for much higher values of $k_{th}b$, since the effective range of atom-atom interactions is much smaller than $\mu$m. Realization of our proposal with cold atoms would extend the existing one-dimensional family of cold-atom spin filters~\cite{Micheli2004,Marchukov2016,Lebrat2019} to the three-dimensional world.

In addition, our results pave the way for investigations of collective scattering from non-atomic 2D crystals that do not allow for spin-flip ($\uparrow \leftrightarrow \downarrow$) transitions at low energies, e.g., systems with a spin gap.  
The CISS effect is a noteworthy phenomenon to study in this regard. In the CISS effect molecules are non-magnetic and possess helical symmetry, which induces chiral effects~\cite{Blum1998}. These properties are not included in our model. 
However, our zero-range model fully incorporates all relevant low-energy information about $\uparrow \to \uparrow$ and $\downarrow \to \downarrow$ scattering processes, and therefore can be used
to estimate the contribution of collective scattering to the CISS effect. 
Zero-range models can describe only the low-energy part of typical electron energies in the CISS experiments, which operate with $0-2\,\mathrm{eV}$ electrons~\cite{Gohler2011,Ray2006}. This places many CISS-related effects beyond our reach. Still, our results can provide an important reference point for future more elaborate theories which will investigate the high-energy regime of CISS.

The CISS experiments can be modelled by a single layer of scatterers with $a_0\neq 0$ and $a_1\neq 0$. Here the parameter $a_0$ should be about the molecular diameter, i.e.,~$1-2\,\mathrm{nm}$~\cite{Aqua2003,Nguyen2019}, and the parameter $a_1$ should be small ($|a_1|\ll |a_0|$) since the spin-orbit coupling is weak for organic molecules. Two important corollaries follow from the analysis of this CISS model. First, the polarization depends weakly on the density of scatterers (it scales as $\sqrt{n}$), which shows that collective interference is important for CISS at low energies. This aligns nicely with the fact that the CISS effect is strong for a wide range of inter-molecular separations $b\sim 1-20\,\mathrm{nm}$~\cite{Aqua2003}. 
Second, multiple scattering for arbitrary values of $a_0,a_1$ and $b$ does not dramatically enhance polarization in comparison to scattering from a single molecule. Therefore, the observed magnitude of the CISS effect can be explained by our model only if the system operates close to the specific parameter regime. This fine tuning is likely since one expects that $|a_0|\simeq b$.
The polarization reversal observed in: (i) molecules embedded in the membrane~\cite{Mishra2013}, and (ii) experiments with a variable temperature~\cite{Eckshtain2016} can be a consequence of that. Indeed, both embedding and temperature denaturation of molecules modify scattering, and hence, the $a_0/b$ ratio, which determines the sign of the polarization coefficient (see Fig.~\ref{fig:TransPolSingleSheet}~{(b)}).

\section{Acknowledgments}
This work has received funding from the European Union's Horizon 2020 research and innovation programme under the Marie Sk\l{}odowska-Curie Grant Agreement No. 754411 (A. G. V. and A. G.). M.L.~acknowledges support by the Austrian Science Fund (FWF), under project No.~P29902-N27, and by the European Research Council (ERC) Starting Grant No.~801770 (ANGULON).

\bibliography{spinpolarizerbib}
\bibliographystyle{apsrev4-1}

\end{document}

\setcounter{figure}{0}
\makeatletter
\renewcommand{\figurename}{Supplementary Figure}
\setcounter{equation}{0} \makeatletter
\renewcommand \theequation{S\@arabic\c@equation}
%\renewcommand \thetable{S\@arabic\c@table}

%\title{Supplementary material for the paper	\protect \\{\bf Spin Filtering in Scattering off Point Magnets}}

\title{Supplementary information for the paper	\protect \\{\bf Filtering Spins by Scattering from a Lattice of Point Magnets}}

\author{Areg Ghazaryan}
\affiliation{IST Austria (Institute of Science and Technology Austria), Am Campus 1, 3400 Klosterneuburg, Austria}

\author{Mikhail Lemeshko}
\affiliation{IST Austria (Institute of Science and Technology Austria), Am Campus 1, 3400 Klosterneuburg, Austria}

\author{Artem G. Volosniev}
\affiliation{IST Austria (Institute of Science and Technology Austria), Am Campus 1, 3400 Klosterneuburg, Austria}
\maketitle
\section*{Supplementary Notes}
%\section{Finite energy solution for scattering off a single crystal}
\subsection*{Supplementary Note 1}
\begin{figure}
\includegraphics[width=12cm]{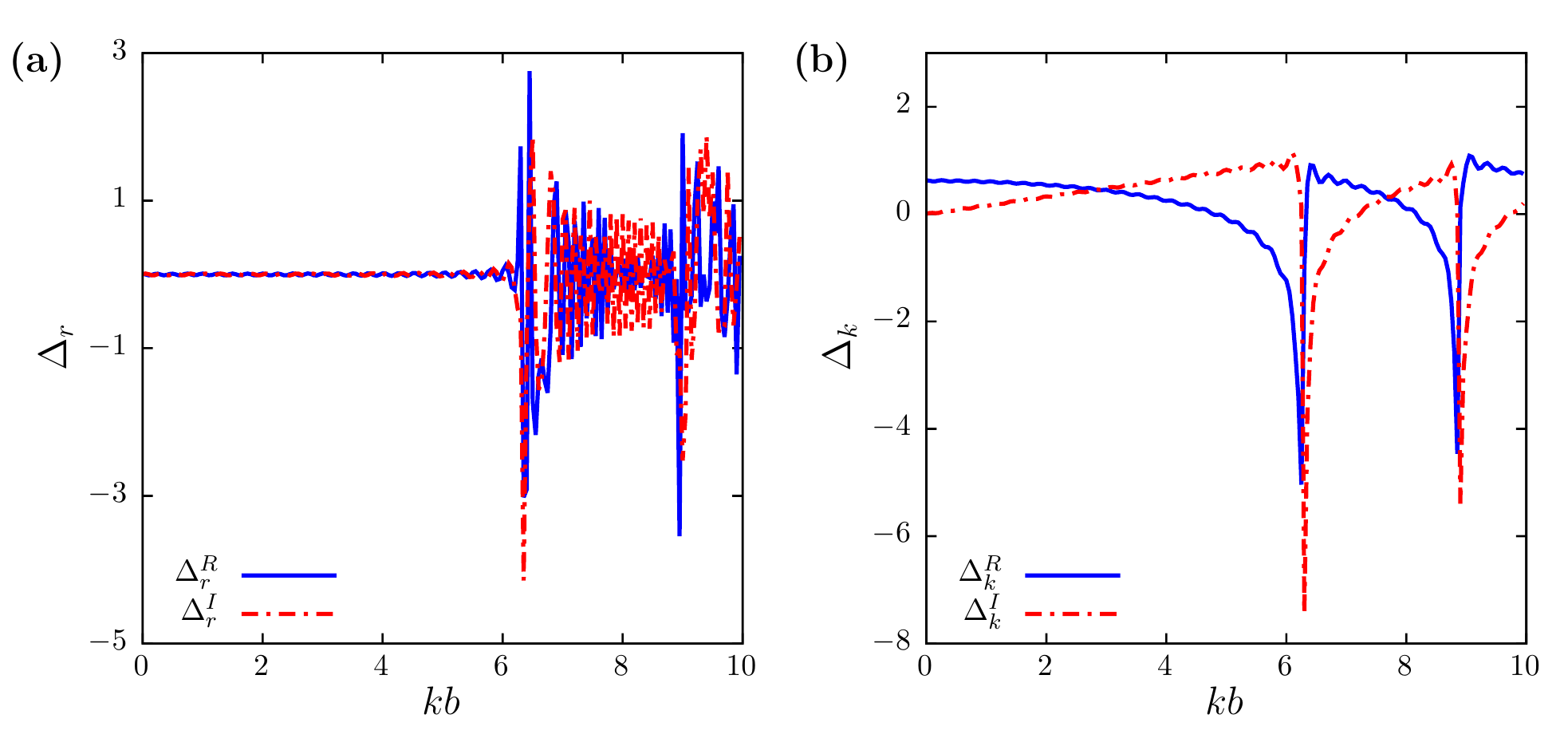}
\caption{\label{fig:DeltakrDep} (a) Dependence of the real ($\Delta^R_r$) and imaginary ($\Delta^I_r$) parts of $\Delta_r$ on the momentum of incoming electrons, $kb$, for $z=40b$. (b) Same as in (a), but for $\Delta_k$. We use $R=400$ in both panels.
}\end{figure}

In this section, we derive the transmission and polarization coefficients for a single layer of point scatterers assuming low but finite energies of incoming electrons. Our starting point is the wave function presented in the main text
\begin{equation}
\Psi\left(\mathbf{r}\right)=e^{ikz}+\lim_{R\to\infty} A_R\sum_{lm}^R \frac{e^{ik\left|\mathbf{r}-\mathbf{a}_{lm}\right|}}{\left|\mathbf{r}-\mathbf{a}_{lm}\right|},
\label{Wavefunct1L}
\end{equation}
where
\begin{equation}
A_R=-\alpha_s\left(1+\alpha_s\sum\limits_{\substack{lm \\ \mathbf{a}_{lm}\neq0}}^R\frac{e^{ik\left|\mathbf{a}_{lm}\right|}}{\left|\mathbf{a}_{lm}\right|}+ik\alpha_s\right)^{-1},
\label{AForm1L}
\end{equation} 
and $R$ defines an upper limit of the sum. As in the main text, to estimate the sums in Eqs.~(\ref{Wavefunct1L}) and~(\ref{AForm1L}), we approximate summations by integrals:
\begin{equation}
\sum^R_{lm}\frac{e^{ik\left|\mathbf{r}-\mathbf{a}_{lm}\right|}}{\left|\mathbf{r}-\mathbf{a}_{lm}\right|}=\frac{2\pi}{ikb^2}\left(e^{ik\sqrt{z^2+b^2R^2}}-e^{ik|z|}\right)-\frac{2\pi}{b}\Delta_r.
\label{FirstSum}
\end{equation}
\begin{equation}
\sum^R\limits_{\substack{lm \\ \mathbf{a}_{lm}\neq0}}\frac{e^{ik\left|\mathbf{a}_{lm}\right|}}{\left|\mathbf{a}_{lm}\right|}=\frac{2\pi}{ikb^2}\left(e^{ikbR}-1\right)-\frac{2\pi}{b}\Delta_k,
\label{SecondSum}
\end{equation}
where the parameters $\Delta_r$ and $\Delta_k$ denote the difference between the exact values of the sums and the integral approximation. For simplicity, we have assumed that $|z|\gg b$ and $\mathbf{r}=z\hat{\mathbf{z}}$. The parameters $\Delta_r$ and $\Delta_k$ depend on the momentum of incoming electrons, $kb$. Supplementary Figure~\ref{fig:DeltakrDep} shows this dependence for $R=400$, although we have checked that the results do not change substantially by changing $R$. For the zero-energy case, $kb=0$, we have $\Delta_r=0$ and $\Delta_k=\Delta_0$ ($\Delta_0\approx0.635$), see the main text. Note that both sums may diverge for $kb\geq 2\pi$ due to constructive interference. We do not discuss this phenomenon here, since the zero-range potential model is applicable only at low energies -- our focus is on $kb\leq 1$. In this energy regime $\Delta_r$ and $\Delta_k$ change weakly, allowing one to describe low-energy scattering using the zero-energy solution considered in the main text. Since $\Delta_r\simeq 0$, from now on we will ignore $\Delta_r$ and only keep $\Delta_k$.  

\begin{figure}
\includegraphics[width=18cm]{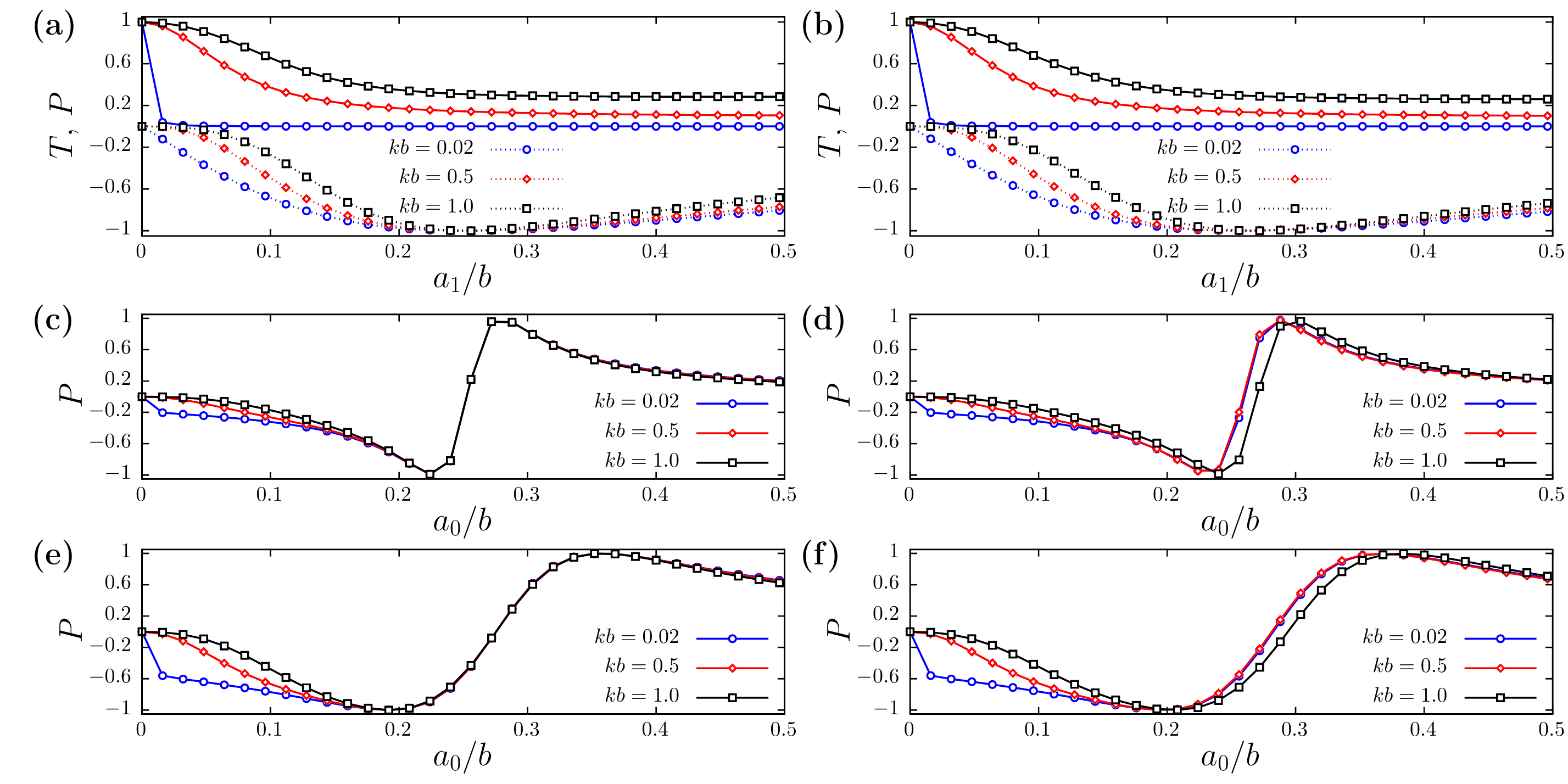}
\caption{\label{fig:TransPolSingleSheetSup} (a, b) Dependence of transmission (points connected by solid lines) and polarization (points connected by dotted lines) on the dimensionless scattering length $a_1/b$ for $a_0=0$.  Dependence of polarization on $a_0/b$ when (c, d) $a_1/a_0=0.1$  and (e, f) $a_1/a_0=0.3$. The left panels (a, c, e) show the results of the zero-range approximation discussed in the main text, while the right panels (b, d, f) present the full solution. The transmission coefficients $T$ is defined as $T=\left(T_\uparrow+T_\downarrow\right)/2$.
}\end{figure}

The $R$-dependent parts of Eqs.~(\ref{FirstSum}) and~(\ref{SecondSum}) do not cancel each other out as in the zero-energy case. Despite that we can ignore them, since an incoming electron beam is a square-integrable wave packet -- the $R$-dependent terms are highly oscillatory and for $R\to \infty$ their integral contribution is negligible. In the limit $|z|\gg b$, the wave function reads
\begin{equation}
\Psi\left(\mathbf{r}\right)\simeq e^{ikz}+\frac{2\pi \alpha_se^{ik|z|}}{ikb^2-2\pi \alpha_s-i2\pi \alpha_s\Delta_k k b-k^2b^2\alpha_s}.
\label{WavefunctFin1L}
\end{equation}     
Comparing this wave function to the outgoing flux in scattering from a 1D Dirac delta potential~\cite{Griffiths2018}, $g_s\delta(z)$:
\begin{equation}
\Psi_{1D}(z)=\frac{2ike^{ikz}}{2ik-g_s},
\end{equation}
we derive a mapping onto a 1D problem, which is similar to the one derived for the zero-energy case. The strengh of the corresponding 1D potential is 
\begin{equation}
g_s=\frac{4\pi \alpha_s}{b\left(b-2\pi \alpha_s\Delta_k+ik\alpha_s b\right)};
\label{ScattAmpl}
\end{equation}
the transmission and reflection coefficients have the form
\begin{align}
T_s&=1-\frac{4\pi^2\alpha_s^2+4\pi kb\alpha_s^2\left(kb-2\pi\Delta_k^I\right)}{\alpha_s^2\left(k^2b^2+2\pi\left(1-kb\Delta_k^I\right)\right)^2+k^2b^2\left(b-2\pi \alpha_s\Delta_k^R\right)^2}, \\
R_s&=\frac{4\pi^2\alpha_s^2}{\alpha_s^2\left(k^2b^2+2\pi\left(1-kb\Delta_k^I\right)\right)^2+k^2b^2\left(b-2\pi \alpha_s\Delta_k^R\right)^2},
\end{align} 
where $\Delta_k^R$ and $\Delta_k^I$ denote the real and imaginary parts of $\Delta_k$. The polarization is determined by $P=\frac{T_\uparrow-T_\downarrow}{T_\uparrow+T_\downarrow}$.
Supplementary Figure~\ref{fig:TransPolSingleSheetSup} compares the transmission and polarization coefficients from above to the ones obtained using the zero-energy approximation, see the main text. We use parameters as in Fig.~2 of the main text, in fact, Supplementary Figs.~\ref{fig:TransPolSingleSheetSup} (b,d,f) shows the results of Fig.~2 of the main text. As is evident from Supplementary Fig.~\ref{fig:TransPolSingleSheetSup}, the results obtained for the zero-energy limit in the main text are very similar to those obtained here. The only considerable difference is the dependence of the specific point, where the layer turns into perfect spin filter, on $k$, which stems from the dependence of $\Delta_k$ on $k$.

\begin{figure}
\includegraphics[width=18cm]{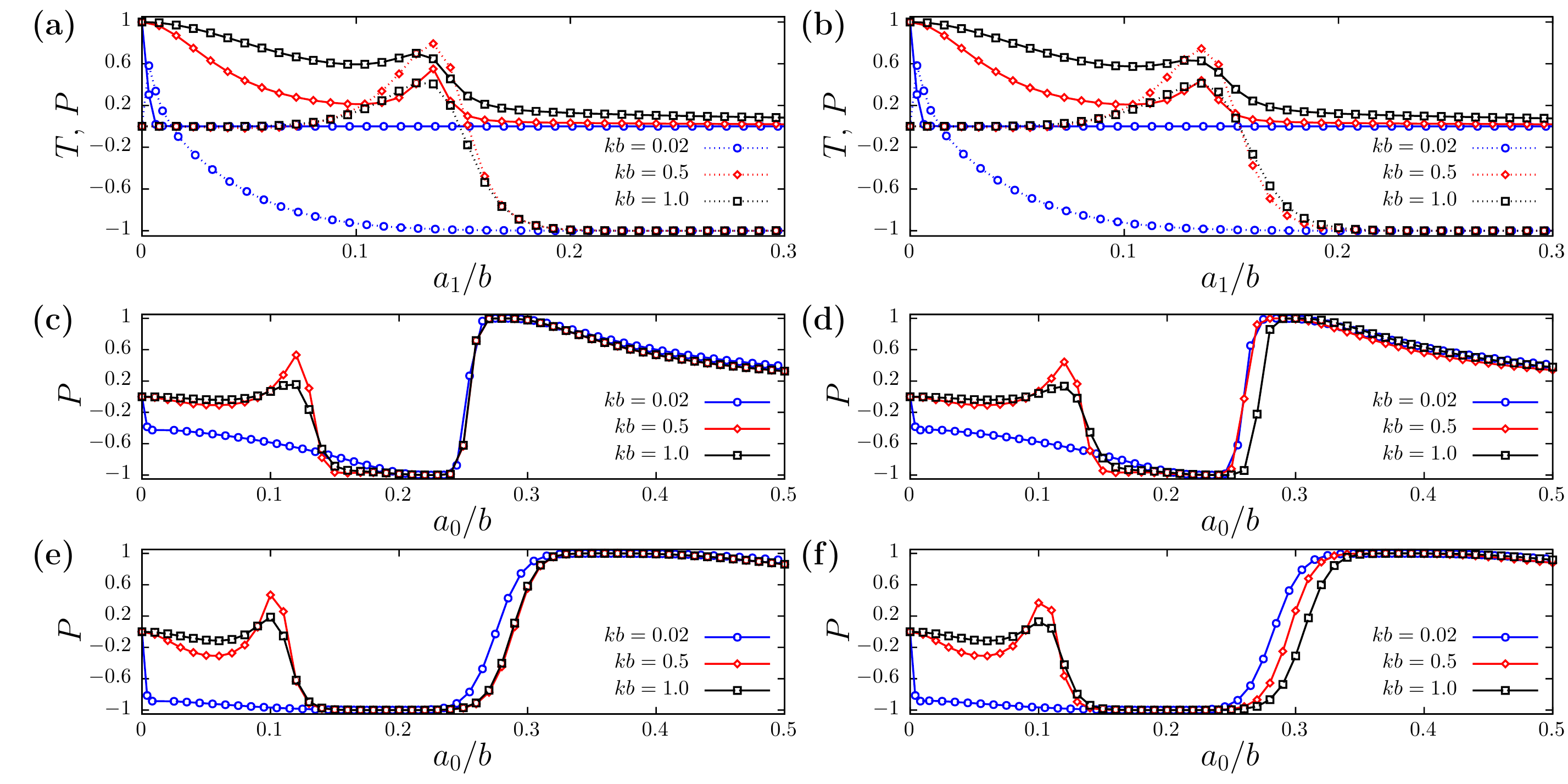}
\caption{\label{fig:TransPolDoubleSheetSup} The same as in Supplementary Figure~\ref{fig:TransPolSingleSheetSup} but for scattering from two layers of point magnets. The separation between the sheets is $L=100b$. 
}\end{figure}

%\section{Finite energy solution for scattering off two parallel crystals}
\subsection*{Supplementary Note 2}
\noindent 
In this section, we consider scattering from two layers separated by distance $L$. Our ansatz for the wave function is
\begin{equation}
\Psi_L\left(\mathbf{r}\right)=e^{ikz}+\lim_{R\to \infty}\sum_iA^i_R\sum_{lm}\frac{e^{ik\left|\mathbf{r}-\mathbf{a}^i_{lm}\right|}}{\left|\mathbf{r}-\mathbf{a}^i_{lm}\right|},
\label{Wavefunct2L}
\end{equation}
where the superscript $i=0,1$ implies that the quantity is related either to the first or the second layers, $\mathbf{a}^0_{lm}=lb\hat{\mathbf{x}}+mb\hat{\mathbf{y}}+0\hat{\mathbf{z}}$ and $\mathbf{a}^1_{lm}=lb\hat{\mathbf{x}}+mb\hat{\mathbf{y}}+L\hat{\mathbf{z}}$. The parameters $A^i_R$ are determined from the boundary conditions
\begin{equation}
\Psi_L\left(\mathbf{r}\rightarrow\mathbf{a}^i_{lm}\right)=s^i_{lm}\left(\frac{1}{\left|\mathbf{r}-\mathbf{a}^i_{lm}\right|}-\frac{1}{\alpha_{s}}\right),
\label{BoundaryConditions2L}
\end{equation}
where, for simplicity, we assume that all scatterers are identical. We obtain from these boundary conditions that
\begin{equation}
A^i_R=\frac{\alpha_s+i\alpha_s^2k+\alpha_s^2S_0-\alpha_s^2S_1e^{(-1)^ikL}}{\alpha_s^2\left(S^2_1-S^2_0\right)-2i\alpha_s^2kS_0-2\alpha_sS_0+\alpha_s^2k^2-2i\alpha_sk-1}.
\end{equation}
Here $S_0$ and $S_1$ are the sums:
$S_i=\sum^R_{lm}e^{ik\left|\mathbf{a}^i_{lm}\right|}/\left|\mathbf{a}^i_{lm}\right|$, hence $S_0$ is the same as (\ref{SecondSum}) and $S_1$ is connected to (\ref{FirstSum}) if $z=-L$ and $L\gg b$. After approximating the sums by integrals [cf.~Eqs~(\ref{FirstSum}) and~(\ref{SecondSum})] and dropping all highly oscillatory terms, we write the wave function in the limit $z\gg L$  as
\begin{equation}
\Psi_{L}\left(\mathbf{r}\right)\simeq\frac{k^2b^2\left(b-2\pi \alpha_s\Delta_k+ik\alpha_sb\right)^2e^{ikz}}{4\pi^2\alpha_s^2e^{2ikL}+\left(2i\pi \alpha_s+kb^2-2k\pi \alpha_sb\Delta_k+ik^2\alpha_sb^2\right)^2}.
\label{WavefunctFin2L}
\end{equation}
This can again be compared with 1D scattering from two identical Dirac delta potentials located at $z=0$ and $z=L$, for which the outgoing flux is \cite{Griffiths2018}
\begin{equation}
\Psi_{1D,L}(z)=\frac{4k^2e^{ikz}}{g_s^2e^{2ikL}+\left(ig_s+2k\right)^2}.
\end{equation}
If we use in this expression the interaction strength from Eq.~(\ref{ScattAmpl}), then we derive Eq.~(\ref{WavefunctFin2L}). Therefore, one can use the mapping derived for a single crystal to describe scattering from multiple crystals, provided that $L\gg b$. The transmission coefficients will have the same form as in the zero-energy case (see the main text), 
\begin{equation}
T_s=\left|\frac{4k^2}{\left[g_s^2e^{2ikL}+\left(ig_s+2k\right)^2\right]}\right|^2,
\end{equation}
with the sole redefinition of $g_s$ according to Eq.~(\ref{ScattAmpl}).
Similar to the single-layer case, the results derived from the zero-energy approximation agree well with the results derived here, see Supplementary Fig.~\ref{fig:TransPolDoubleSheetSup}. The only change is a weak energy dependence of the specific point.

\def\bibsection{}
\bigbreak
\section*{Supplementary References} 
\bigbreak
\bibliography{spinpolarizerbib}
\bibliographystyle{apsrev4-1}